\documentclass[final,english]{bullsrsl}[2022/06/15]

\usepackage[utf8]{inputenc}
\usepackage[T1]{fontenc}
\usepackage{graphicx}
\usepackage{natbib}

\begin{document}
\title{Efficient Identification and Analysis of Space debris Trails in Time-Domain Astronomy.  }

\author[affil={1,10}, corresponding]{Sarvesh Kumar}{Yadav}
\author[affil={3,1}]{Bhavya}{Ailawadhi}
\author[affil={4,5}]{Talat}{Akhunov}
\author[affil={1,2}]{Naveen}{Dukiya}
\author[affil={6}]{Sara}{Filali}
\author[affil={7,8}]{Paul}{Hickson}
\author[affil={1}]{Priyanshi}{Kumari}
\author[affil={1}]{Gokul Singh}{Mehra}
\author[affil={1}]{Kuntal}{Misra}
\author[affil={9}]{Vibhore}{Negi}
\author[affil={1,10}]{Kumar}{Pranshu}
\author[affil={1,6}]{Jean}{Surdej}
\author[affil={1,2}]{Monalisa}{Dubey}

\affiliation[1]{Aryabhatta Research Institute of observational sciencES (ARIES), Manora Peak, Nainital, 263001, India}
\affiliation[2]{Department of Applied Physics, Mahatma Jyotiba Phule Rohilkhand University, Bareilly, 243006, India}
\affiliation[3]{Physical Research Laboratory, Ahmedabad 380009, Gujarat, India}
\affiliation[4]{National University of Uzbekistan, Department of Astronomy and Astrophysics, 100174 Tashkent, Uzbekistan}
\affiliation[5]{Ulugh Beg Astronomical Institute of the Uzbek Academy of Sciences, Astronomicheskaya 33, 100052 Tashkent, Uzbekistan}
\affiliation[6]{Institute of Astrophysics and Geophysics, Li\'ege University, All\'ee du 6 Ao\^ut 19c, 4000 Li\'ege, Belgium}
\affiliation[7]{Department of Physics and Astronomy, The University of British Columbia, 6224 Agricultural Road, Vancouver, BC V6T 1Z1, Canada}
\affiliation[8]{Outer Space Institute, The University of British Columbia, 325-6224 Agricultural Road, Vancouver, BC V6T 1Z1, Canada}
\affiliation[9]{Kavli Institute for Astronomy and Astrophysics, Peking University, Beijing 100871, China}
\affiliation[10]{Department of Applied Optics and Photonics, University of Calcutta, Kolkata, 700106, India}

\correspondance{sarveshkyadav.5201@gmail.com}

\maketitle

\begin{abstract}

This paper presents an automated end-to-end pipeline to detect and characterize artificial satellites and space debris streaks. As the exponential rise of orbital debris poses an escalating risk to global space assets, providing scalable, real-time tools is a critical contribution to the global Space Situational Awareness (SSA) network. The architecture utilizes a multi-stage geometric approach for raw image segmentation and high-precision morphological characterization. Experimental results demonstrate robust performance in real-world survey environments, successfully recovering streak profiles with an operational detection limit for features with integrated $SNR \ge 86$ and lengths $\ge 98$ arcseconds. This unsupervised framework strengthens global space sustainability efforts by monitoring RSO populations.

\end{abstract}

\vspace{1em}

\section{Introduction}

The rapid growth of Resident Space Objects (RSOs)---comprising active satellites, spent rocket stages, and orbital debris---poses a challenge to Space Situational Awareness (SSA) and ground-based optical astronomy \citep{ESA_Debris_2025}. In long-exposure astronomical imaging, RSOs manifesting as non-sidereal trailed lines, or ``streaks,'' contaminate observational data and interfere with the photometric and astrometric analysis of celestial objects. Given the massive data volumes produced by modern high-cadence sky surveys, manual identification of these features is computationally impractical and non-scalable, necessitating the development of robust, automated detection frameworks \citep{automated_detection}.

The International Liquid Mirror Telescope (ILMT), located at Devasthal, India, serves as a dedicated facility for deep-sky zenithal survey. Utilizing a 4-meter liquid primary mirror, the ILMT operates in Time-Delay Integration (TDI) mode, where the CCD charge transfer is synchronized with the sidereal rate \citep{ILMT}. Consequently, any object moving with a non-sidereal angular velocity produces a linear streak across the detector. While initial RSO identification at the ILMT relied on visual inspection \citep{manual_detection}, transitioning to nightly high-cadence operations requires an autonomous pipeline capable of real-time detection and characterization.

Conventional methods for linear feature extraction in high-noise astronomical images include Matched Filtering and the Radon Transform. Although Matched Filtering is effective for specific morphologies, it lacks the flexibility to handle the varying widths and orientations of RSOs without significant template overhead \citep{matched_filtering_radon}. The Radon Transform, while mathematically rigorous, is often computationally expensive for large-format CCD frames \citep{matched_filtering_radon, Hickson_2018}. In contrast, the Hough Transform provides an efficient, unsupervised mechanism for detecting collinear pixel distributions. By mapping image-space coordinates into a parametric $(\rho, \theta)$ accumulator space, the algorithm remains resilient against low signal-to-noise ratios (SNR) and fragmented streak profiles, making it particularly well-suited for the unsupervised detection of orbital debris \citep{zhu2024sstdstripelikespacetarget}.

Several studies have used the Hough Transform for streak detection in astronomical images. Vananti et al. \citep{vananti2015} used it alongside the Radon Transform to pick out faint debris tracks from noisy CCD frames, Diprima et al. \citep{diprima2018} combined it with morphological filtering for fast, automated detection, and more recent tools like ASTA \citep{stoppa2024} blend deep learning with a Probabilistic Hough refinement step. Most of these approaches, however, rely on external reference data, pre-labeled training sets, or a single-stage detection step that limits how much morphological detail can be recovered from each streak. The pipeline described here takes a different approach --- it uses a two-stage Hough scheme, first locating the streak with a Probabilistic Hough Transform and then measuring its full extent with a Standard Hough Transform, all built on top of LoG-based edge detection and profile fitting tuned to how streaks actually appear in ILMT frames. The result is a fully automated pipeline that needs no manual checks or training data, and goes beyond simple detection to characterize each streak's length, width, and SNR.

We detail this end-to-end architectural design below, from raw image preprocessing to the optimization of the parametric voting mechanism, and evaluate its performance using representative detections from ILMT observational cycles.


\section{Methodology}
The proposed detection and characterization pipeline is designed to manage the high volume of ILMT data using a modular, multi-stage processing architecture. The pipeline is designed to maximize computational efficiency while ensuring robust localization and characterization of non-sidereal streaks. By partitioning the workflow into discrete functional phases, the pipeline facilitates high-speed processing and provides a comprehensive set of morphological and photometric parameters for each detected streak.

\begin{figure}
    
    \centering
    \includegraphics[width=1\linewidth]{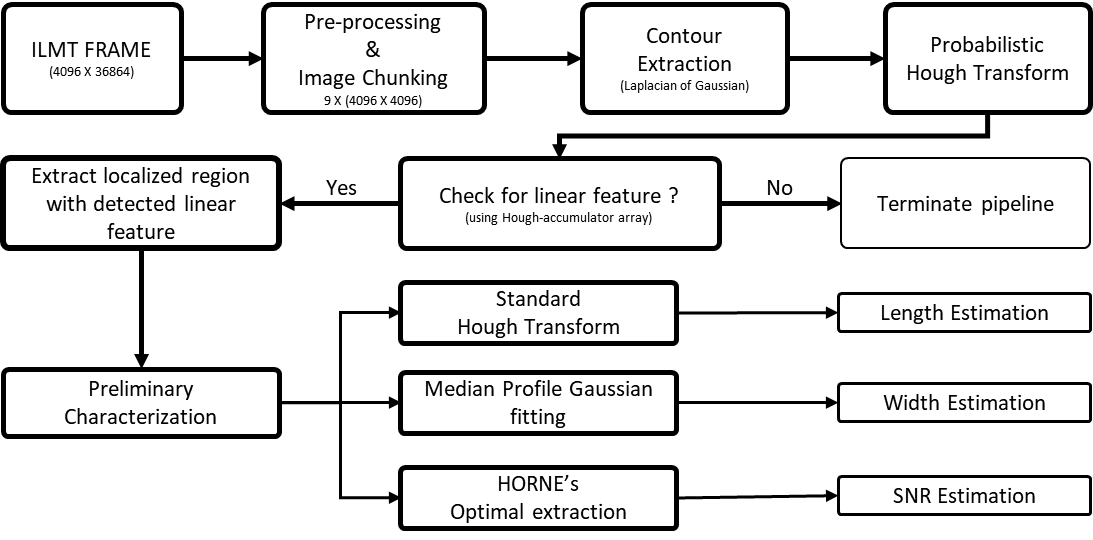}
    \begin{minipage}{12cm}
    \caption{Schematic representation of the automated detection and characterization pipeline for the ILMT frames, illustrating the progression from initial image segmentation (chunking) to final feature parameter estimation.}
    \label{fig:Pipeline}
    
    \end{minipage}

\end{figure}

As illustrated in Figure~\ref{fig:Pipeline}, the operational workflow initiates with image segmentation (chunking) to optimize memory management during large-frame processing. This is followed by a hierarchical detection scheme that integrates a Laplacian of Gaussian (LoG) operator for edge enhancement and a dual-stage Hough Transform for linear feature isolation. Potential candidates are validated using the Hough accumulator array. Subsequently, the pipeline enters a characterization phase where the Standard Hough Transform (SHT), Gaussian fitting, and Horne's Optimal Extraction are utilized to quantify the streak's length, width, and signal-to-noise ratio (SNR). The specific technical implementation of these stages is detailed in the following subsections.

\subsection{Image Pre-processing and chunking: }

\begin{figure}
    \centering
    \includegraphics[width=1\linewidth]{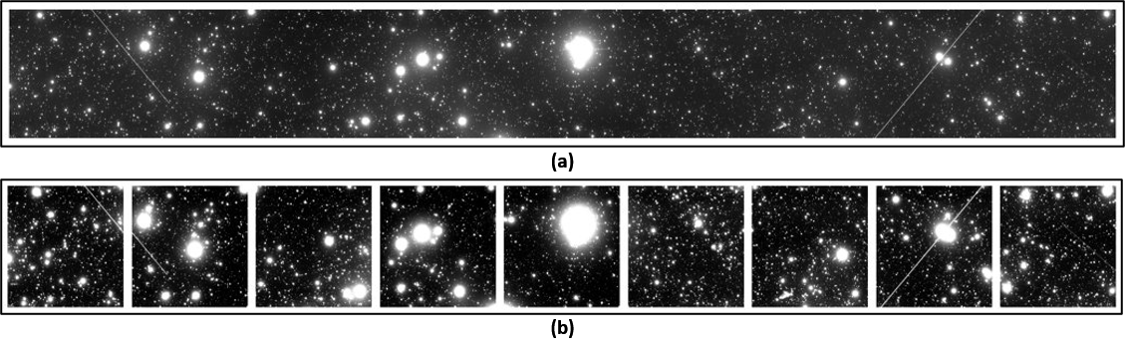}
    \begin{minipage}{12cm}
    \caption{Multi-stage image pre-processing workflow: (a) Original calibrated $4k \times 36k$ ILMT long-format frame; (b) Processed $4k \times 4k$ sub-frames (chunks) following background suppression and dynamic range compression to isolate potential linear features from the stochastic background noise.}
    \label{fig:stage_1}
    
    \end{minipage}
\end{figure}

The ILMT generates large-format CCD frames with dimensions of $4,096 \times 36,864$ pixels. Following initial astrometric and photometric calibration, each frame occupies approximately 576 MB of memory \citep{ILMT_DATA}. A representative sample of this long-format observational data is displayed in Figure~\ref{fig:stage_1}a.

To suppress high-frequency artifacts and noise, a $3 \times 3$ median filter is applied to the raw frames. This non-linear filtering stage is utilized to suppress stochastic noise and residual cosmic-ray hits while preserving the sharp edges of potential streaks \citep{manual_detection}. Given the high memory consumption of the full-frame data, the image is segmented into nine discrete $4,096 \times 4,096$ pixel sub-frames, or ``chunks,'' to facilitate parallelized processing and optimize memory management (Figure~\ref{fig:stage_1}b).

The subsequent phase focuses on optimizing the pixel intensity distribution to enhance the detectability of low signal-to-noise ratio (SNR) features. As depicted in Figure~\ref{fig:stage_1}(b), the pipeline implements dynamic range compression by clipping pixel intensities at a $5\sigma$ threshold above the local median. This prevents saturation from bright sidereal sources and mitigates the risk of blooming artifacts obscuring faint RSO signals. Concurrently, a floor-subtraction is performed by resetting all pixel values below a $Median - 1\sigma$ threshold to zero. This dual-thresholding effectively suppresses stochastic background fluctuations, thereby increasing the contrast of potential faint-streak candidates relative to the background.

\subsection{Streak Detection and Localization: }

\begin{figure}
    \centering
    \includegraphics[width=0.98\linewidth]{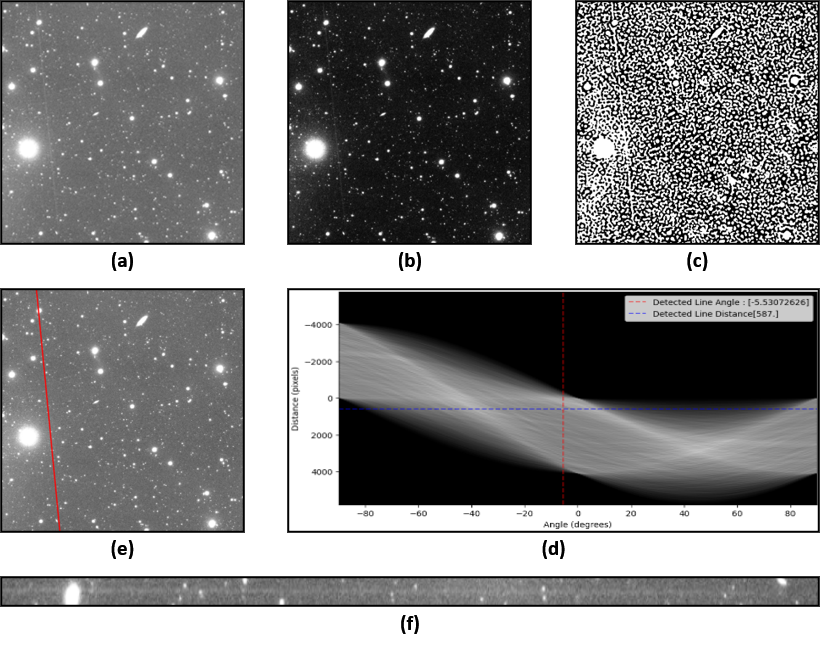}

    \begin{minipage}{12cm}
    \caption{The streak detection and localization sequence: (a) raw chunk; (b) processed frame with enhanced contrast; (c) LoG contour extraction; (d) Hough-accumulator array showing the peak detection; (e) localized line projection; and (f) final 100-pixel wide region of detected streak. }
    \label{fig:streak_localization}
    
    \end{minipage}    
\end{figure}

The detection sequence initiates with the $4,096 \times 4,096$ pixel chunk (Figure~\ref{fig:streak_localization}a). Because faint orbital streaks are frequently obscured by the ambient stellar background and Poisson noise, the pipeline uses thresholding and background suppression techniques detailed in the previous section. The resulting processed frame (Figure~\ref{fig:streak_localization}b) enhances the contrast of the streak candidates relative to the surrounding stellar population, facilitating subsequent feature extraction.

To isolate potential linear candidates, the pipeline employs the LoG operator for contour extraction (Figure~\ref{fig:streak_localization}c). The LoG is a second-order derivative filter that performs Gaussian smoothing to mitigate noise sensitivity, followed by the Laplacian operator to identify regions of rapid intensity \citep{LoG}. The LoG of an image $I(x,y)$ is mathematically defined as $L(x,y) = \Delta [G_{\sigma}(x,y) * I(x,y)]$.

where $G_{\sigma}$ denotes a Gaussian kernel with standard deviation $\sigma$ and the symbol `$*$' represents the discrete 2D convolution operation. This operator is specifically optimized to highlight the edges of linear structures while suppressing uniform background regions, thereby generating a binary mask for parametric analysis.

The extracted binary contours are subsequently processed using the Probabilistic Hough Transform (PHT) to localize the streak in parametric space (Figure~\ref{fig:streak_localization}d). The PHT utilizes a random subset of points to reduce computational overhead while maintaining detection accuracy \citep{prob_hough}. The transform maps each point $(x,y)$ from the image domain to a sinusoidal curve in the $(\rho, \theta)$ parameter space according to the normal equation of a line, $\rho = x \cos \theta + y \sin \theta$.

Collinear points in the image space present as intersecting curves at a discrete $(\rho, \theta)$ coordinate in the accumulator array. By identifying local maxima within this array---representing the highest point density or ``votes''---the pipeline determines the distance ($\rho$) and angular orientation ($\theta$) of the linear feature.

Following parametric determination, the detected line is back-projected onto the image coordinates (Figure~\ref{fig:streak_localization}e). This localization allows the pipeline to isolate a specific Region of Interest (ROI). In the final stage of this process (Figure~\ref{fig:streak_localization}f), the localized streak is cropped into a 100-pixel wide ROI. This segmentation minimizes extraneous stellar noise and provides a constrained data frame for the precise characterization of the streak's length, width, and signal-to-noise ratio (SNR).

\subsection{Preliminary Streak Characterization:}
The characterization phase derives morphological and photometric parameters from the 100-pixel wide ROI using a sequence of specialized estimation modules.

\begin{figure}[h]
    \centering
    \includegraphics[width=1\linewidth]{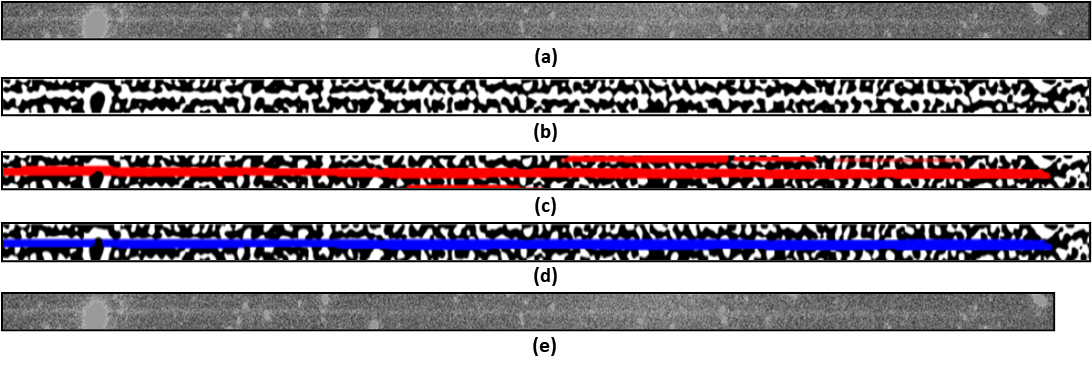}
    \begin{minipage}{12cm}
        \caption{The length estimation sequence: (a) masked ROI; (b) LoG contour; (c) initial SHT segments; (d) primary streak cluster; and (e) the final length-cropped ROI.}
    \label{fig:length_estimate}
    \end{minipage}
\end{figure}

\textbf{Length Estimation:} While the PHT is optimized for localization, it frequently yields fragmented segments. To recover the full linear extent, the SHT is applied to the LoG contour map of the ROI (Figure~\ref{fig:length_estimate}b). The SHT exhaustively evaluates every pixel, allowing the pipeline to aggregate disjoint segments (Figure~\ref{fig:length_estimate}c) into a primary cluster based on shared $(\rho, \theta)$ intercepts \citep{standard_hough}. Merging these collinear components defines the true endpoints and total length, resulting in a refined ROI (Figure~\ref{fig:length_estimate}e) for subsequent analysis.

\begin{figure}[htbp]
    \centering
    \begin{minipage}{0.6\linewidth}
        \centering
        \includegraphics[width=\linewidth]{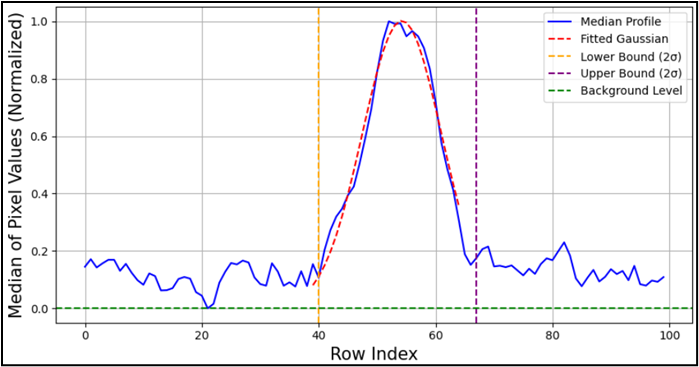}
    \end{minipage}
    \hfill
    \begin{minipage}{0.35\linewidth}
        \caption{Width estimation via Gaussian fitting: The blue line represents the median cross-sectional profile of the ROI, while the dashed red line shows the fitted Gaussian within a 25-pixel ($\sim$8~arcseconds) window. The orange and violet dashed lines define the $4\sigma$ effective width of the streak (Estimated Width = 27 pixels).}
        \label{fig:width}
    \end{minipage}
\end{figure}

\textbf{Width Estimation:} The transverse width is determined by fitting a Gaussian model to the median cross-sectional intensity profile (Figure~\ref{fig:width}). To mitigate stochastic noise and artificial profile broadening, the fitting is restricted to a 25-pixel window centered on the peak intensity. The effective width is defined by a $4\sigma$ interval, comprising $2\sigma$ on either side of the fitted mean, to encompass approximately 95\% of the signal flux while minimizing background contamination.

\textbf{SNR Estimation:} Photometric precision is maximized through Horne's Optimal Extraction \citep{HORNE}, which weights pixels by their expected spatial profile. The extracted flux ($F$) is defined as:

\begin{equation}
    F = \frac{\sum_{i} (D_i P_i / \sigma_{i}^{2})}{\sum_{i} (P_{i}^{2} / \sigma_{i}^{2})}
\end{equation}

where $D_i$ is the pixel data, $P_i$ is the normalized profile model, and $\sigma_{i}^{2}$ is the local variance. By prioritizing the high-intensity core over the distribution wings, this method significantly enhances the integrated SNR relative to standard aperture photometry. For the sample ROI, this implementation achieved an integrated SNR of around 139.

\section{Results and Conclusions}

The pipeline was installed and executed on a system with an Intel Core i7-14700K processor (20 cores), 16\,GB of RAM, running Ubuntu 24.04.3 LTS. On this system, processing a single $4k \times 4k$ chunk, including all preprocessing and Hough Transform-based detection, takes around 5 seconds, which means a full ILMT frame (divided into nine such chunks) is scanned in under a minute. If a streak candidate is found, the follow-up characterization and orbital cross-matching add roughly another minute, assuming that the relevant TLE catalogs are already available.

\begin{figure}[ht]
    \centering
    \includegraphics[width= 1\linewidth]{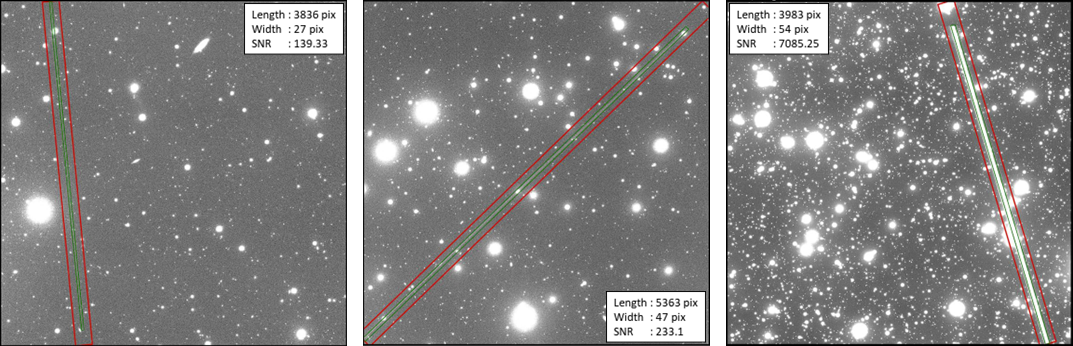}

    \begin{minipage}{12cm}
    \caption{Representative streak detections illustrating the pipeline's performance across different signal levels. Red boxes indicate the PHT-localized ROI; green boxes denote the final morphological characterization (length and width). The integrated SNR is observed to increase progressively from left to right.}
    \label{fig:results_gallery}
    
    \end{minipage}
\end{figure}

The automated detection and characterization pipeline was evaluated using a diverse set of observational frames from the ILMT survey. Figure~\ref{fig:results_gallery} illustrates three representative samples of detected Resident Space Objects (RSOs) with varying signal intensities. From left to right, the samples show increasing streak intensity, and all were successfully recovered by the pipeline. The outer red rectangular boundaries represent the ROI initially localized via PHT. The inner green boundaries represent the precise longitudinal (length) and transverse (width) morphology derived from the SHT and Gaussian fitting modules, respectively.

\begin{figure}[htbp]
    \centering
    \begin{minipage}{0.6\linewidth}
        \centering
        \includegraphics[width=\linewidth]{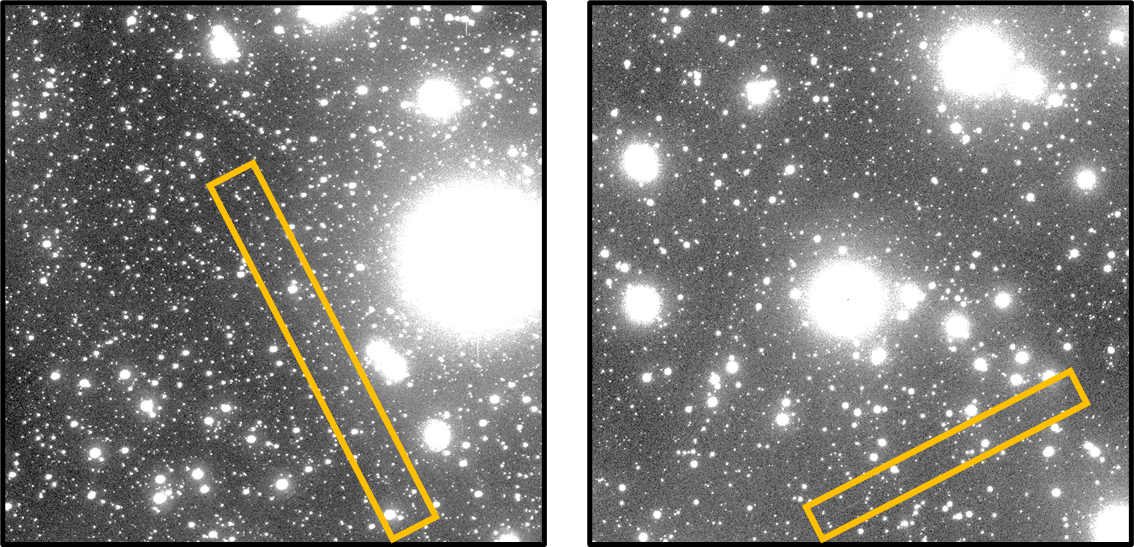}
    \end{minipage}
    \hfill
    \begin{minipage}{0.35\linewidth}
        \caption{Samples of undetected streaks (missed detections) due to low signal intensity. The yellow boxes indicate the positions of faint streaks identified through manual inspection that fell below the pipeline's automated detection threshold.}
        \label{fig:missed_detections}
    \end{minipage}
\end{figure}

Quantitative analysis confirms that the pipeline effectively extracts the structural profiles and the SNR without manual intervention. However, a rigorous comparative analysis against manually identified streaks reveals specific operational constraints of the Hough-based approach. The detection sensitivity is bounded by a minimum threshold; streaks with an integrated SNR below 86 were consistently missed by the automated framework. Furthermore, the voting mechanism in the Hough-accumulator array requires a minimum density of collinear points to distinguish a feature from stochastic noise. Consequently, linear features with a longitudinal extent of less than approximately 300 pixels fail to generate a sufficiently high peak in the parametric space for reliable detection.

Examples of these limitations are shown in Figure~\ref{fig:missed_detections}, where extremely faint streaks (highlighted in yellow) were undetected by the pipeline due to their low contrast. Despite these parametric constraints, the developed architecture provides a robust, high-speed solution for the majority of RSO populations encountered in nightly ILMT operations. By automating the transition from raw data ingestion to photometric characterization, this work establishes an efficient framework for real-time Space Situational Awareness and provides a scalable foundation for future deep-learning integration (Yadav et al.; in preparation) to lower the current SNR detection limits.

\begin{acknowledgments}
The 4-m International Liquid Mirror Telescope (ILMT) project results from a collaboration between the Institute of Astrophysics and Geophysics (University of Li`{e}ge, Belgium), the Universities of British Columbia, Laval, Montreal, Toronto, Victoria, and York University, and Aryabhatta Research Institute of Observational Sciences (ARIES, India). The team acknowledges the contributions of ARIES's technical and scientific staff. JS wishes to thank Service Public Wallonie, F.R.S.-FNRS (Belgium), and the University of Li`{e}ge, Belgium, for funding the construction of the ILMT. PH acknowledges financial support from the Natural Sciences and Engineering Research Council of Canada, RGPIN-2019-04369. PH, SF, and JS thank ARIES for hospitality during their visits to Devasthal. We sincerely acknowledge Anna Pospiezalska for her significant contributions to ILMT functioning. MD acknowledges the Innovation in Science Pursuit for Inspired Research (INSPIRE) fellowship award (DST/INSPIRE Fellowship/2020/IF200251) for this work. KP acknowledges the support from the Erasmus+ Programme of the European Union for a research visit to the Institute of Astrophysics and Geophysics, University of Li`ege, Belgium. TA thanks the Ministry of Higher Education, Science, and Innovations of Uzbekistan (grant FZ-20200929344). JS and KM acknowledge the assistance received from the ANRF SERB-726 VAJRA Faculty Scheme, India. KM acknowledges the support from the BRICS grant DST/ICD/BRICS/Call-5/CoNMuTraMO/2023 (G) funded by the DST, India. We acknowledge the Belgo-Indian Network for Astronomy and astrophysics (BINA) consortium approved by the International Division, Department of Science and Technology (DST, Govt. of India; DST/INT/BELG/P-09/2017) and the Belgian Federal Science Policy Office (BELSPO, Govt. of Belgium; BL/33/IN12), for allowing us to present our work in the form of a paper in the proceedings of the fourth BINA workshop.

\end{acknowledgments}

\begin{furtherinformation}

\begin{orcids}
\orcid{0009-0002-2355-5626} {Sarvesh Kumar } {Yadav}
\orcid{0000-0003-1637-267X} {Kuntal } {Misra}
\orcid{0000-0002-7005-1976} {Jean} {Surdej}

\end{orcids}

\begin{authorcontributions}
All authors contributed significantly to the work presented in this paper.
\end{authorcontributions}

\begin{conflictsofinterest}
The authors declare no conflict of interest.
\end{conflictsofinterest}
\end{furtherinformation}

\bibliographystyle{bullsrsl-en}

\bibliography{references}

\end{document}